# *Ab-initio* modeling of electrolyte molecule Ethylene Carbonate decomposition reaction on $Li(Ni,Mn,Co)O_2$ cathode surface

Shenzhen Xu[1], Guangfu Luo[1], Ryan Jacobs[1], Shuyu Fang[2], Mahesh K. Mahanthappa[3], Robert J. Hamers[2], Dane Morgan[1*]

*[1]Department of Materials Science and Engineering, [2]Department of Chemistry*

*University of Wisconsin – Madison, Madison, WI, USA*

*[3]Chemical Engineering and Materials Science, University of Minnesota, Minneapolis, MN*

*Email: ddmorgan@wisc.edu

## Abstract

Electrolyte decomposition reactions on Li-ion battery electrodes contribute to the formation of solid electrolyte interphase (SEI) layers. These SEI layers are one of the known causes for the loss in battery voltage and capacity over repeated charge/discharge cycles. In this work, density functional theory (DFT)-based *ab-initio* calculations are applied to study the initial steps of the decomposition of the organic electrolyte component ethylene carbonate (EC) on the $(10\bar{1}4)$ surface of a layered $Li(Ni_x,Mn_y,Co_{1-x-y})O_2$ (NMC) cathode crystal, which is commonly used in commercial Li-ion batteries. The effects on the EC reaction pathway due to dissolved $Li^+$ ions in the electrolyte solution and different NMC cathode surface terminations containing adsorbed hydroxyl -OH or fluorine –F species are explicitly considered. We predict a very fast chemical reaction consisting of an EC ring-opening process on the bare cathode surface, the rate of which is independent of the battery operation voltage. This EC ring-opening reaction is unavoidable once the cathode material contacts with the electrolyte because this process is purely chemical rather than electrochemical in nature. The –OH and –F adsorbed species display a passivation effect on the surface against the reaction with EC, but the extent is limited except for the case of –OH bonded to a surface transition metal atom. Our work implies that the possible rate-limiting steps of the electrolyte molecule decomposition are the reactions on the decomposed organic products on the cathode surface rather than on the bare cathode surface.

## 1. Introduction

A combination of transition metal oxide cathodes with organic solvent-based electrolytes is commonly used in current commercial rechargeable Li-ion batteries. Due to the high operation voltage of Li-ion batteries, the organic molecules in the electrolyte are likely to react with the oxidizing cathode materials at the cathode-electrolyte interface.[1-2] These side reactions lead to irreversible capacity loss and voltage fade over numerous charging/discharging cycles,[3] which has led to a number of experimental and theoretical efforts to develop approaches to protect the cathode materials [4-12]. Understanding the mechanisms controlling the interfacial reactions between the cathode material and the electrolyte molecules is of great interest for improving the chemical stability between the cathode and electrolyte.

In this work, we have focused on the initial step of the electrolyte molecule ethylene carbonate (EC) decomposition reaction on the $Li(Ni_x,Mn_y,Co_{1-x-y})O_2$ (NMC) cathode surface. EC solvent and NMC are commonly used as electrolyte components and cathode material in Li-ion batteries, respectively. The bulk reaction between EC and NMC is thermodynamically favorable. This is the driving force for the EC decomposition reaction on the cathode material. However, in the real system of cathode particle coexisting with EC molecules, the reaction between EC and NMC is constrained because of the formation of a solid electrolyte interphase (SEI) around the cathode material as a protecting layer. The SEI layer can form on both the cathode/electrolyte and the anode/electrolyte interfaces due to electrolyte molecule decomposition reactions on these electrodes. At the anode/electrolyte interface, the formation of the SEI layer has been studied extensively in previous experimental and simulation works.[1-2, 13-15] However, the possible electrolyte decomposition reactions on the surface of oxidizing cathode materials are less well understood. Previous experimental works have already shown evidence of electrolyte decomposition products on the surface of cathode materials (e.g., LiF,[16] carbon dioxide,[17-18] organic radicals[19] and unidentified polymers[20-21]), yet the mechanism of formation for these SEI species on the cathode remains unclear. In particular, it is not known whether the SEI formation on the cathode is mainly caused by the direct electrolyte molecule decomposition reaction on the cathode surface or by the collection of decomposed products which have migrated from the anode [22]. In addition, if the cathode-electrolyte reaction does occur on the surface, it is not known whether such a reaction is predominantly chemical or electrochemical, and therefore sensitive in nature to the cell voltage.

Previous density functional theory (DFT)-based simulation studies have provided useful insight into the kinetics of EC decomposition on Li-ion battery cathode surfaces.[23, 24] Here, we summarize the main findings of these previous works, which we will compare with our current findings in detail in Section 3.2. Previous theoretical work by Leung[23] studied the EC decomposition reaction on the spinel $LiMn_2O_4$ cathode (100) surface under ultrahigh vacuum (UHV) conditions. In that work, a fast EC bond breaking process on the bare $LiMn_2O_4$ cathode (100) surface with a barrier of 0.24 eV was predicted. Another recently published theoretical paper by Tebbe et al.[24] modeled the $LiCoO_2$ ($LiCoO_2$ is also a layered-structure material that is isomorphous with the NMC material in this study) cathode surface in the presence of EC molecules. They predicted that the EC ring-opening process is facilitated either by $PF_5$ (one of the decomposition products of $LiPF_6$) or an extra EC molecule from the electrolyte solution. Their calculated reaction barriers to decompose EC are higher than 1 eV for both of these cases. In this work, we extend these previous studies in multiple ways, including simulating a new materials system (NMC), a more realistic electrolyte environment, and varied states of hydroxylation and fluorination of the surface. We show that both the latter two aspects have a large impact on the decomposition pathways and barriers. In Section 3.2, we compare our modeling work to the above two works (Leung, 2012; Tebbe, et al., 2016) and discuss the differences and implications in detail.

## 2. Methods

## 2.1 DFT calculations

To find the transition state between the intermediates of the EC decomposition reaction and to calculate the corresponding reaction barrier, we have used the climbing-image nudged elastic band (CI-NEB) method[25] implemented in the Vienna *ab initio* Simulation Package (VASP)[26] DFT calculation software. The projector augmented wave method (PAW)[27] was used for the pseudopotentials of all atoms. The valence electron configurations were $2s^2 2p^4$ for O, $2s^2 2p^2$ for C, $1s^1$ for H, $1s^2 2s^1$ for Li, $3d^8 4s^1$ for Co, $3p^6 3d^6 4s^1$ for Mn, and $3p^6 3d^9 4s^1$ for Ni. The generalized gradient approximation (GGA) exchange-correlation functional of the Perdew-Wang (PW-91)[28] version is used with the Hubbard U correction (GGA+U)[29] applied to the transition metals. The U values for Ni, Mn and Co were 6.37 eV, 4.84 eV, and 5.14 eV, respectively, which were obtained from previous simulation work[30] where the U values of Ni, Mn and Co were determined by fitting to the experimental Li intercalation voltages of the layered-structure $LiNiO_2$, $LiMnO_2$, and $LiCoO_2$ cathode materials, respectively. In our slab calculations, the two bottom layers of atoms were fixed to the relaxed bulk structure, and the two top layers of atoms in the surface slab were relaxed (see Fig. 1(a)). In the recently work of Tebbe, et al. (2016) [24] where they studied the $LiCoO_2$ $(10\bar{1}4)$ surface (which is similar to our case), the top two layers were relaxed as well. They reported in their paper that the convergence of the reaction barrier value is within 8.5meV if they increase the three-layer slab with the top two layers being relaxed to a four-layer slab with the top three layers being relaxed. This test from the work of Tebbe, et al. (2016) [24] validates that our surface slab thickness (four-layer slab) is sufficient and supports the surface slab relaxation method (top two layers being relaxed) used in our model. A 400 eV plane wave energy cutoff and a Γ-centered k-point mesh were used to obtain the reaction barrier energy converged within 20 meV.

## 2.2 NMC cathode surface modeling

The atomistic structure of the simulated NMC $(10\bar{1}4)$ surface slab with an EC molecule is shown in Fig. 1(a). The chemical formula of the NMC slab is $Li_{48}(Ni_{24}Mn_{14}Co_{10})O_{96}$, the relative atomic ratios [Ni]:[Mn]:[Co] = 5:3:2 as in commonly used experimental and commercial NMC cathodes.[31-32] The Ni, Mn and Co atoms were randomly distributed amongst the transition metal lattice sites in the surface slab. The cleavage surface in our simulation is the edge-plane $(10\bar{1}4)$ surface, which is believed to be dominant surface for Li transport in and out of the cathode material, and therefore of particular interest to consider when exploring SEI formation.[33] The $(10\bar{1}4)$ surface also has a lower surface energy compared to the other two possible edge-plane $(10\bar{1}0)$ and $(11\bar{2}0)$ surfaces,[33] making the $(10\bar{1}4)$ surface the most stable edge-plane surface. The in-plane lattice parameters of the surface slab (*x*- and *y*-directions shown in Fig. 1(a)) were obtained from the fully relaxed bulk $Li_{48}(Ni_{24}Mn_{14}Co_{10})O_{96}$ supercell, which are 11.76 Å and 17.77 Å, respectively. The length of the supercell along the *z*-direction (perpendicular to the terminating $(10\bar{1}4)$ surface) is 38 Å. The thickness of the surface slab (containing 192 atoms) is 6.3 Å. The length of the vacuum along the z-direction is 31.7 Å to ensure there is enough vacuum space to avoid spurious interactions between the periodic images of the surface slabs. In our DFT simulation, the NMC cathode surface slab is in a ferromagnetic state.

The surface EC reactions would ideally be modeled in systems with multiple layers of electrolyte molecules. However, such a large and complex simulation is not computationally tractable and we therefore approximate the electrolyte environment. Specifically, we consider the EC bond breaking in two environments. First, we consider just an isolated EC molecule with no electrolyte, or an Ultra-High Vacuum (UHV) condition. The UHV condition is a useful simple baseline on which to build our more complex model of an electrolyte solution. Our second environment includes the influence of $Li^+$ and other nearby EC molecules on the EC breaking, and represents an approximation to the electrolyte environment. The interaction of the EC molecule with $Li^+$ is expected to be the most important influence of the surrounding electrolyte on the EC cathode surface reaction barriers because, in the real electrolyte organic solution (EC/DEC (ethylene carbonate/diethyl carbonate), EC/DMC (ethylene carbonate/dimethyl carbonate), etc.), the dissolved $Li^+$ ion is usually coordinated by up to 4 EC molecules.[34] In our model, the $Li^+$ is surrounded with 4 EC molecules as shown in Fig. 1(b). The positive charge state of the Li was obtained by extracting one electron from the supercell (the positive Li charge state was verified by the Bader charge method[35], which resulted in a charge of +0.9, and can be approximately considered as $Li^+$).

It should be noted here that our modeled solution environment is subject to the periodic boundary condition of our unit cell and this may cause some quantitative change to our calculated reaction barrier result. One source of this surface-size induced influence is the interaction between the EC that we studied and its nearest image EC from the solution environment (the 3 EC molecules on top of the $Li^+$). However, we think the corresponding impact on the exact value of the calculated EC bond breaking reaction barrier should be very limited. We find that the distance between the EC molecule (which goes through the C-O bond breaking process) and its nearest image EC molecule from the solution environment is about 10.1Å, with the closest atoms between the molecules still 7.8Å apart. This distance is large enough that we expect essentially no interaction between the molecules. However, to be sure, we have done a test DFT calculation to estimate the interaction between the two EC molecules which are 10Å apart from each other (with the closest atoms between the molecules 7.8Å apart). Our calculation shows that if the two EC molecules are moved towards each other by 0.4Å (in the $C_C$-$O_E$ bond breaking process, the $O_E$ moves away from the $C_C$ atom by about 0.4 Å and forms a bond with a surface transition metal atom). The total energy of the two-EC system is only increased by 2meV. It indicates that the interaction between the two EC molecules which are 10Å away from each other is negligible.

Another possible effect of the periodic boundary conditions is that it can force an unphysical structure at the interface. However, we are not attempting to construct an accurate surface electrolyte model but instead a local environment around the EC that is realistic. The $Li^+$-$O_C$ distance is a key intermolecular bond length which can be used to check if the $Li^+$ with 4 EC structure in our model can well represent the local environment in a real electrolyte solution. Two previous simulation papers reported the calculated $Li^+$-$O_C$ distance in an explicit electrolyte solution model ($LiPF_6$ solute in EC solvent). In the work of Tasaki (2002) [36], the average $Li^+$-$O_C$ distance is 2.06Å. In the work of Wang et al. (2001) [13], the average $Li^+$-$O_C$ distance is 1.97Å. In our model, the average $Li^+$-$O_C$ distance is 2.07Å. So our calculated $Li^+$-$O_C$ distance matches well with the results from previous explicit electrolyte solution models. This result indicates that our modeled local solvent environment can approximate that in an explicit solution model. Overall we think the surface size issue will not have a significant impact on our calculated EC ring-opening reaction barrier.

In addition to the electrolyte environment at the surface, three possible surface terminations are considered in our model. As a baseline, we first consider the bare cleaved surface with no additional surface species. However, due to exposure to the environment, *e.g.*, exposure to air during synthesis or the electrolyte solution after battery assembly, the cathode surface may no longer be bare. In this work we have therefore also considered the cathode surface terminated with a hydroxyl group –OH and a fluorine atom –F, which we believe are the most likely surface terminating species given the cathode environments.[37-39] A possible source of the hydroxyl group –OH is the $H_2O$ from the air, which may interact with the cathode surface during the cathode synthesis and the battery cell assembly process,[37] or $H_2O$ present in trace amounts in the electrolyte. The –F may come from the $PF_6^-$ salt and/or other fluoride species, e.g., HF, which has been detected in the electrolyte solution.[40-41] To include the hydroxyl group and the fluorine atom in our atomistic model of the NMC surface, it is important to consider how much hydroxylation and the fluorination is expected. Hydroxylation is a ubiquitous phenomenon on oxide surfaces and results from the interaction of the oxide surface with water:[37]

$$(M_{surface} - O_{surface}) + H_2O \leftrightarrow (O_{surface} - H_{(ads)}) + (M_{surface} - OH_{(ads)}) \qquad \text{Eq. 1}$$

During construction of Li-ion batteries, the particles comprising the Li battery cathode always undergo prolonged heating (temperature range is about 600 K-900 K) in vacuum or inert gas conditions to remove the $H_2O$ from the cathode [24, 37]. Despite this prolonged heat treatment, experimental observations have shown that the hydroxyl termination of cathode surfaces still persists. For example, prolonged heating of $LiCoO_2$ for 4 h at 623 K under ultrahigh vacuum conditions gives approximately 1/3 hydroxyl group monolayer surface coverage.[37] A higher heating temperature (up to about 900 K) and longer heating period may lower the hydroxyl surface coverage. In addition, reaction with water in the electrolyte might yield some hydroxyl coverage. In general, although a significant hydroxyl coverage of the surface is possible, we expect significantly less than a monolayer of coverage due to the aggressive efforts to remove all water from the cathode and battery systems. For the surface coverage of the fluorine atom, as the main source of –F is the electrolyte solution, the –F coverage on the cathode surface should be nearly zero prior to contact with the electrolyte. Right after the NMC cathode material contacts with the electrolyte, the EC ring-opening process occurs extremely quickly on the bare NMC cathode surface ($E_{\text{barrier}}$ is only 17meV) as shown in Results Section 3.3 (*vide infra*). Therefore, although we do not know the exact mechanism and extent of the NMC surface fluorination, we assume that the surface chemical environment should have a quite low –F coverage at the moment the EC molecule is going through the ring-opening process on the cathode surface. Based on the above discussion about the –OH and –F surface coverage, we expected that the coverage of both –OH and –F is well below a monolayer, which suggests that these functionalities will be relatively isolated on the surface. We therefore consider the influence of just a single surface –OH or –F species on the decomposition path. Furthermore, as the decomposition pathway is very local, we expect the most important effect of a surface species to be when it covers a site used in the decomposition reaction, and we focus on this type of configuration in our barrier studies.

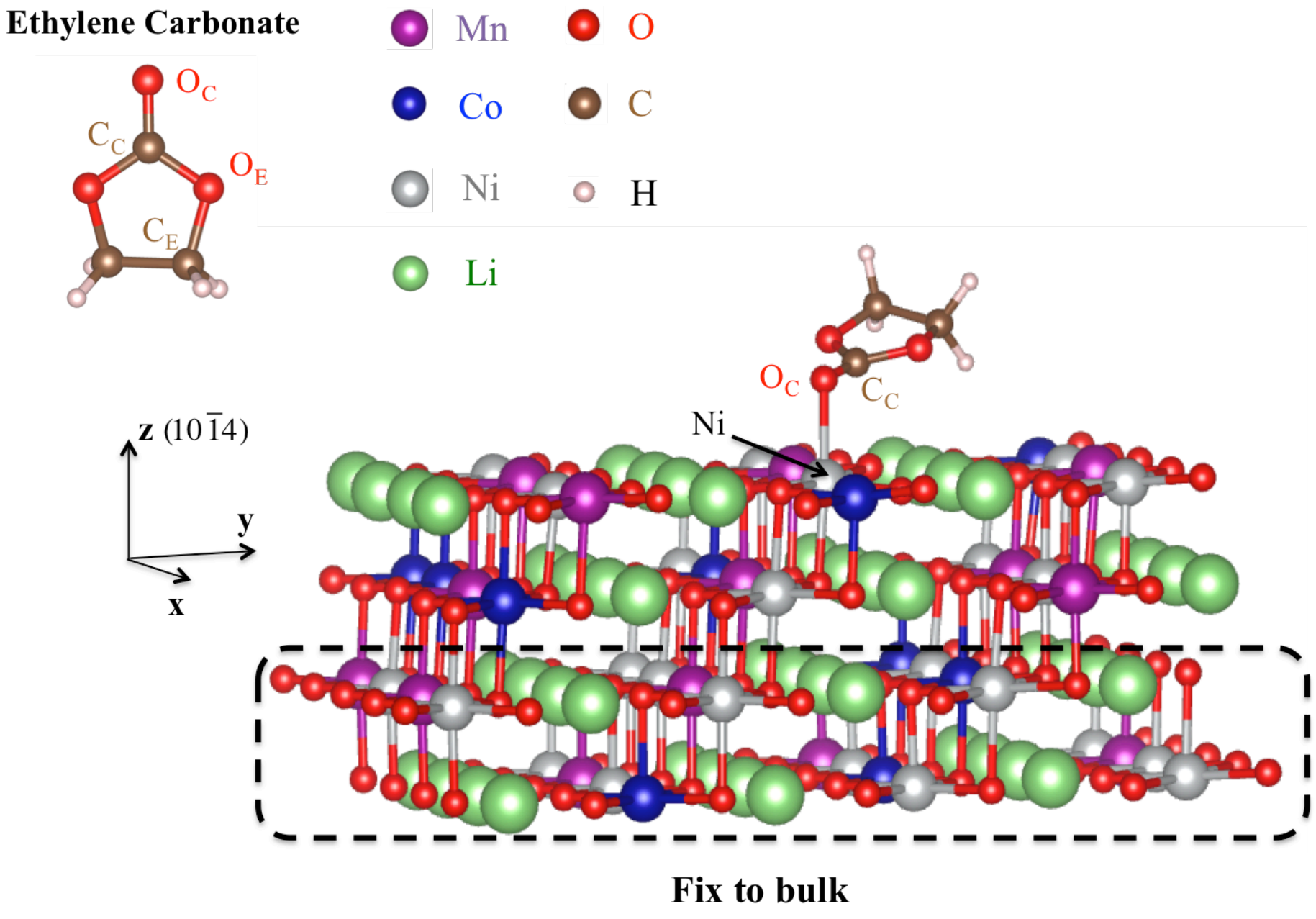

(a)

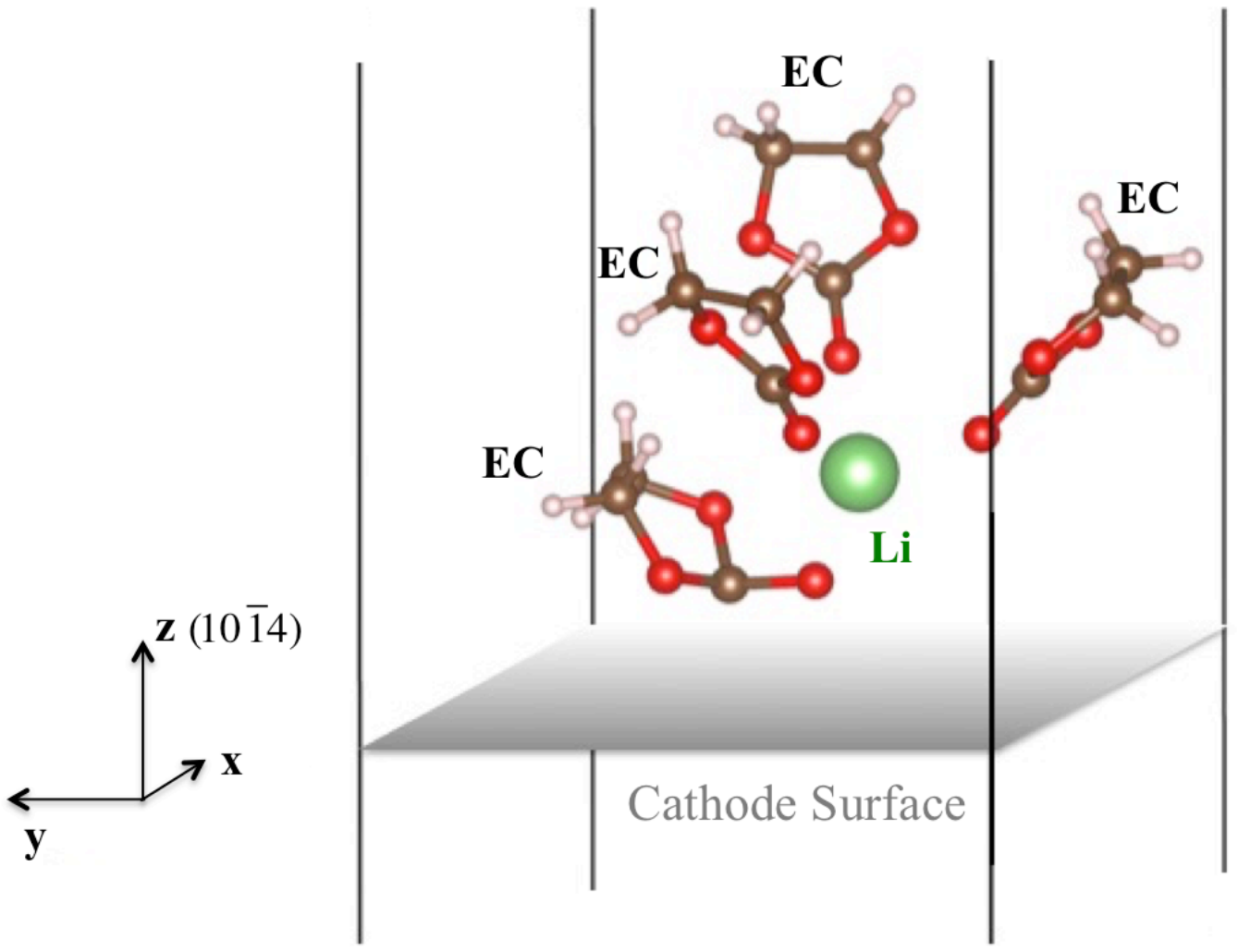

(b)

**Fig. 1**: Structural illustrations of our modeled surface system. (a) The atomistic structure of the $Li_{48}(Ni_{24}Mn_{14}Co_{10})O_{96}$ NMC cathode surface slab with an EC molecule bonded to a transition metal atom on the surface. The x-y plane is the $(10\bar{1}4)$ surface. The EC structure is shown at the top. The names for the specific atoms in the EC molecule are carbonyl oxygen ($O_C$), carbonyl carbon ($C_C$), ethylene oxygen ($O_E$) and ethylene carbon ($C_E$). (b) The atomistic configuration of the $Li^+$ ion (in the electrolyte) coordinated with 4 EC molecules. The EC molecule closest to the cathode surface (the relative position is underneath the other 3 EC molecules in the plot) is the one that reacts on the cathode surface.

## 3. Results

### 3.1 Key reaction steps of the EC ring-opening process

In the previous simulation work of Leung,[23] it was shown that the initial EC decomposition reaction step on the cathode surface involved breaking the $C_C$-$O_E$ (carbonyl carbon - ethylene oxygen) bond, and the rate-limiting step of the $C_C$-$O_E$ bond breaking process was the $C_C$ atom forming a bond with a surface oxygen atom $O_{surface}$. We first considered the same (meta)stable intermediates as identified Leung's work, and Fig. 2 shows the fully relaxed configurations of these intermediate steps of the EC bond breaking process on the NMC cathode surface. The first state (Fig. 2(a)) is the initial state where an EC molecule approaches the cathode surface and the carbonyl oxygen atom $O_C$ (in EC) bonds with a transition metal (TM) atom on the surface (Ni in our case). After the initial state shown in Fig. 2(a), the configuration evolves to the second state (the tetrahedral intermediate) in Fig. 2(b) where the carbonyl carbon atom $C_C$ bonds with an oxygen atom on the cathode surface $O_{surface}$. Lastly, the third state in Fig. 2(c) shows the bond breaking between the $C_C$ and the $O_E$ in the EC molecule while the $C_C$ and the $O_E$ are still connected to the cathode surface through a nearby TM (Mn in our case). The $O_C$ remains bonded to the Ni throughout the three steps.

It should be noted here that in our modeling work, we only consider the case where the EC molecule is initially bonded to a surface Ni and splits to also bond to a surface Mn, and do not consider any other TM atom combinations with Mn, Ni, and Co. The focus on one TM pair greatly simplifies the study and we believe that changing the TM element species would only have a minor influence on the reaction barrier of the $C_C$-$O_E$ bond breaking process. The reasons we expect only a weak influence of TM species are, first, that changing the TM element species would not alter the EC bond-breaking pathway. As shown in our following NEB calculations of the reaction pathways, the main chemical function of the TM atom on the cathode surface appears to be attracting and bonding with the oxygen atoms in the EC molecule. This attraction of the oxygen atoms to the TM centers on the surface is expected to occur regardless of the specific TM species. The exact bond length and the bond stability may vary slightly for the different TM elements, but this variation is not essential to our model of the $C_C$-$O_E$ bond breaking process in EC and does not qualitatively change the physical implications of our calculated reaction barriers. A second reason we consider just one TM is that, from the atomistic configurations of the intermediate steps shown in Fig. 2, we can see that the activation barriers mainly depend on the process of the $C_C$ atom forming a bond with a surface oxygen atom ($O_{surface}$) and the intrinsic process of the $C_C$ atom breaking the bond with the $O_E$ atom in EC. This EC ring-opening process does not include the TM species bonded to the EC, so we believe that the exact TM element species in the local surface structure will not have a significant effect on the barrier. Third and finally, as shown in the following Results Section 3.3, Section 3.4 and the Discussion Section 4, our present studies have already identified a decomposition pathway that is very fast, demonstrating that the EC $C_C$-$O_E$ bond breaking process cannot be the rate-limiting step in the complete EC decomposition reaction on the cathode. While other TMs might give reaction rates that are somewhat faster or slower, such results would not alter the prediction that other rate-liming steps must dominate and such results are not expected to be particularly informative. To supplement the above analysis, we also performed a test calculation to check the reaction barrier change when the surface transition metal species involved in the bond breaking process is changed. The testing case is the bare NMC surface in a local electrolyte solution environment (with the existence of $Li^+$ coordinated with 4 EC molecules). The surface Mn atom which is bonded to the EC $O_E$ atom after the ring-opening process (as shown in Figure 2 (c)) is switched with a surface Co atom. So the $O_E$ atom will bond to the surface Co atom after the EC ring-opening process. The new reaction barrier is 24meV, which is very close to the reaction barrier value of 17meV in the original case (Figure 3(b)). In summary, based on the above arguments and our test calculation, the TM species is not expected to have a significant impact on our mechanistic understanding or predicted rates for EC $C_C$-$O_E$ bond breaking processes on the NMC cathode surface, and only initial $C_C$ bonding to Ni, with final $O_E$ bonding to Mn, is considered.

In previous theoretical work [42] of EC decomposition processes on the $LiMn_2O_4$ [111] surface, it was reported that a step of proton transfer from the EC molecule to the cathode happens before the $C_C$-$O_E$ bond-breaking step. While in the previous theoretical work [23] of EC decomposition process on the $LiMn_2O_4$ [100] surface, the proton transfer step happens after the $C_C$-$O_E$ bond-breaking step. We notice that the difference of the surface structures ([100] vs. [111]) is the key reason leading to the step order change ($C_C$-$O_E$ breaking vs. proton transfer). More specifically, as the transition metal Mn atoms are exposed at the [100] surface, the EC $O_E$ atom can easily bond with the Mn atom at the surface. However, for the [111] surface, there are no transition metal atoms exposed at the surface. The $O_E$ is bonded to a surface Li atom rather than a transition metal atom after the $C_C$-$O_E$ bond breaking. Another difference is that the EC $C_C$ forms a bond with a surface oxygen atom before the $C_C$-$O_E$ bond breaking in the [100] surface case, while in the [111] surface case there is no interaction between the EC $C_C$ and the cathode surface oxygen atom. All these differences may lead to the step order change of the $C_C$-$O_E$ breaking vs. the proton transfer. In our model of the NMC $(10\bar{1}4)$ surface, transition metal atoms are exposed on the surface as well as the surface oxygen atoms, which is more similar to the $LiMn_2O_4$ [100] surface case. Therefore we think the $C_C$-$O_E$ bond breaking process is likely to happen prior to the proton transfer in our modeling system (consistent with the work of Leung (2012) [23]). Another previous theoretical work [43] also considered the proton transfer process as the first step of the EC decomposition reaction on the cathode surface. The system that they studied was a fully-delithiated $(Ni,Mn)_2O_4$ cathode material, which is also different from our modeling case. As discussed in the works [42-43], for the highly-delithiated cathode material, it is easier to go through the proton transfer process first. In our modeling case, we look at fully lithiated NMC cathode surface, which is very different to the condition in the work [43]. A more comparable theoretical work to our modeling case is the work of Tebbe et al., (2016) [24] where they studied the fully lithiated $LiCoO_2$ $(10\bar{1}4)$ surface. The ring-opening process is the first step to happen in their work, which supports our argument that the proton transfer step is likely to happen after the ring-opening process in our fully-lithiated NMC $(10\bar{1}4)$ surface. As a summary, we acknowledge that the proton transfer (from EC to cathode) is one of the possible steps in the EC decomposition reaction, and the reaction rate of the proton transfer may be faster if the cathode material is more delithiated. However, in our modeling system (fully-lithiated NMC $(10\bar{1}4)$ surface), we believe this proton transfer step will follow after the EC $C_C$-$O_E$ bond-breaking process. Therefore we only focus on the $C_C$-$O_E$ breaking reaction as the initial step of EC decomposition in this work.

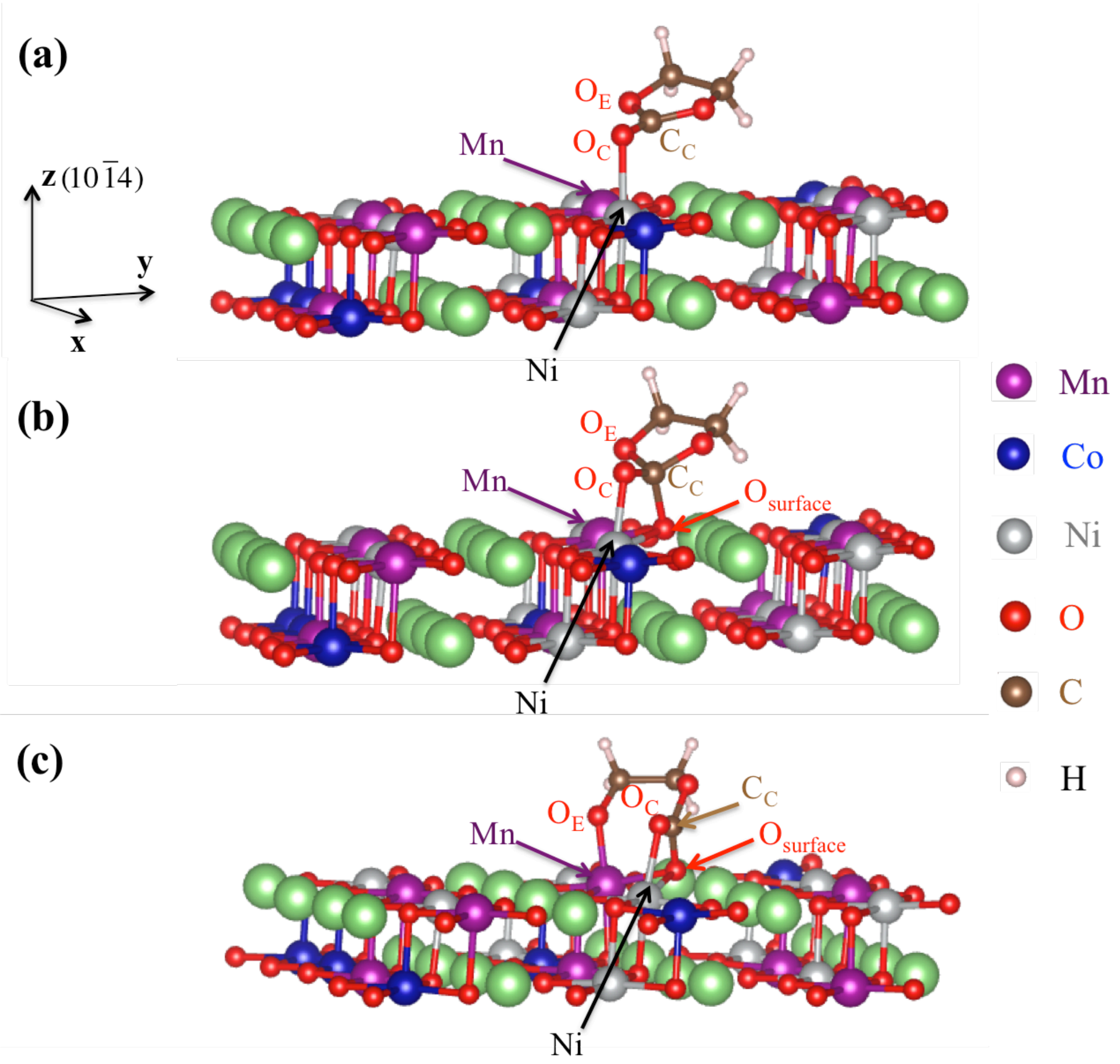

**Fig. 2**: Configurations of the three key intermediate steps, (a)-(c), in the $C_C$-$O_E$ bond breaking process. The x-y plane is the $(10\bar{1}4)$ surface.

### 3.2 Reaction energy barriers of the EC ring-opening process under UHV conditions

After obtaining the atomistic configurations of the three (meta)stable intermediate steps shown in Fig. 2, we performed CI-NEB calculations to obtain the energy barriers between these intermediate steps. In Fig. 3(a), the structures with the image #0, #2 and #4 correspond to the three intermediate steps (a), (b) and (c) in Fig. 2, respectively. The complete EC ring-opening process can be considered as two steps. The first step is the formation of $C_C$-$O_{surface}$ (from intermediate (a) to (b)). The corresponding barrier is 290meV. The second step is the actual $C_C$-$O_E$ bond breaking. The corresponding barrier is 150meV. So the rate-limiting reaction barrier of the EC ring-opening process is 290 meV under the UHV condition. The transition state configuration consists of the carbonyl carbon atom $C_C$ from the EC approaching the surface oxygen atom $O_{surface}$, but still not bonded to the $O_{surface}$.

Here, we provide physical comparisons between our reaction pathway and energy barrier value for EC decomposition on the $(10\bar{1}4)$ NMC cathode surface under UHV conditions with previous simulations.[23-24] These comparisons not only validate our current modeling work, but also reveal new insight in the EC decomposition process on cathode surfaces. In Leung's simulation work[23] of EC $C_C$-$O_E$ bond breaking steps on the spinel $LiMn_2O_4$ cathode (100) surface, the reaction pathway is qualitatively consistent with that shown in Fig. 3(a). In addition, their calculated energy barrier of 240 meV under UHV conditions also matches well with our result of 290 meV. In comparison with another recently published theoretical study of EC decomposition reaction on the $LiCoO_2$ cathode $(10\bar{1}4)$ surface by Tebbe, et al.,[24] our reaction pathway is fundamentally different. In their modeling work,[24] the configuration of the initial state when the EC molecule is bonded to the cathode surface (comparable to the state with the image #0 in Fig. 3(a)) is different from ours. Their initial state has the carbonyl oxygen $O_C$ (in EC) bonded to one of the transition metal sites (for example, the Ni site in our surface structure), and one of the ethylene oxygen $O_E$ bonded to a neighboring transition metal site (for example, the Mn site in our surface structure). Please refer to the Fig. S4 in our Supporting Information (SI) Section 2 for the detailed information about the atomistic structure of the initial state in the work of Tebbe, et al.[24]. For our NMC, we compared the energy between the configuration of the EC adsorption at two neighboring transition metal reported in the work of Tebbe, et al.[24] and the configurations of the intermediate states #0, #2 and #4 shown in Fig. 3(a) of the present work. The energy of the bonding configurations involving two neighboring transition metal centers is 40 meV lower than the energy of state #0, 95 meV higher than the energy of state #2 and 210 meV higher than the energy of state #4 in our NMC cathode surface case. Moreover, the reported energy barriers of the EC bond breaking (ethylene carbon and ethylene oxygen bond $C_E$-$O_E$) processes in the work of Tebbe, et al.[24] are all larger than 1 eV, while the calculated reaction energy barrier in our modeled pathway is only 290 meV. Therefore, aside from the predicted reaction pathway reported by Tebbe, et al.[24], the reaction pathway studied in this work is also a kinetically and thermodynamically feasible mechanism for the EC molecule bond breaking process as the initial step in it complete decomposition reaction. Furthermore, the mechanism in this work is predicted to be significantly faster than that identified by Tebbe, et al., although our studies focus on different cathode systems. The main mechanism missing in the work of Tebbe, et al. is the $C_C$ (in EC) forming a bond with a surface oxygen $O_{surface}$, which weakens the $C_C$-$O_E$ bond in EC and contributes to the possibility of EC $C_C$-$O_E$ bond breaking in the next step. This mechanism is also discussed in Leung's previous simulation work.[23]

Here, we discuss the charge transfer from the EC molecule to the cathode and the change of the magnetic moment and the charge of the transition metal atoms in the cathode before and after the $C_C$-$O_E$ bond breaking process. For the charge transfer analysis, we use Bader charge method [35] to calculate the charge on each atom. We find that there is about 0.56*e* transfer from the EC molecule to the cathode slab after the $C_C$-$O_E$ bond breaking. For the magnetic moment change of the transition metal, we find that only the surface Mn atom (which is bonded with the $O_E$ after the $C_C$-$O_E$ bond breaking) has a significant magnetic moment change, from 3.9 to 3.4, after the $C_C$-$O_E$ bond breaking. This change qualitatively corresponds to a valence state change of Mn from 3+ to 4+. This valence state change is consistent with the coordination number change of Mn (from 5 coordinated O to 6 coordinated O) after the $C_C$-$O_E$ bond breaking. The bader charge change for this surface Mn is -0.14*e*. We also find that the charge of the cathode surface oxygen atom, which is bonded to the $C_C$ (in EC molecule), increases by about 0.6*e* after the $C_C$-$O_E$ bond breaking, indicating that this surface oxygen atom gains most of the electron charge from the EC molecule.

### 3.3 Reaction energy barriers of the EC ring-opening process in an electrolyte environment

Next, we have studied the effect of the electrolyte solution environment, here consisting of $Li^+$ coordinated by 4 EC molecules, on the EC decomposition reaction barrier. We again examine the EC $C_C$-$O_E$ bond breaking process, but now with a EC-coordinated $Li^+$ near the cathode surface, as shown in Fig. 1(b). Fig. 3(b) shows a reaction pathway for EC decomposition in an electrolyte environment similar to the one above for EC decomposition under UHV conditions (Fig. 3(a)). In the presence of $Li^+$, the initial state of the EC molecule, with its $O_C$ bonded to the surface Ni atom and the $Li^+$ ion, spontaneously evolves to the next intermediate state where its $C_C$ forms a bond with the surface oxygen atom $O_{surface}$, with the $O_C$ still bonded to the surface Ni atom and the $Li^+$ ion. This result means that there is no barrier for the $C_C$-$O_{surface}$ bond formation step (the first step discussed in Section 3.2, from intermediate (a) to (b) in Figure 2). Then, the system must overcome a small energy barrier of 17 meV to break the $C_C$-$O_E$ bond (the second step discussed in Section 3.2, from intermediate (b) to (c) in Figure 2). The transition state of this $C_C$-$O_E$ bond breaking process is the image with the image #1 in Fig. 3(b). This transition state corresponds to the moment when the $C_C$ atom just breaks the bond with the $O_E$ atom in the EC molecule. By comparing the reaction energy barriers between the UHV conditions of Fig. 3(a) and electrolyte conditions of Fig. 3(b), it is clear the $Li^+$ and its coordinated EC molecules lower the rate-limiting reaction barrier significantly, from 290 meV to 17 meV. The key physical difference between the UHV and electrolyte conditions is the step where the $C_C$ atom in the EC molecule approaches to the surface oxygen atom to form a bond. This step is the transition state in the UHV condition case without $Li^+$, contributing to a 290 meV barrier. However, when $Li^+$ and its coordinated EC molecules are introduced to the system, the barrier disappears and the $C_C$ atom spontaneously bonds to the surface oxygen atom. The resulting rate-limiting step of 17 meV is due instead to the $C_C$-$O_E$ bond breaking process.

A qualitative explanation for the spontaneous $C_C$-$O_{surface}$ bond formation is that the electrophilic $Li^+$ cation serves as a Lewis acid that binds the ethylene carbonate and activates its carbonyl for nucleophilic attack by the surface oxygen atom. Our DFT calculation shows that the electron charge of the $C_C$ atom decreases by 0.2|e| upon carbonate binding to the $Li^+$ ion. Thus, the $Li^+$ inductively draws electron density away from the carbonyl functionality, rendering the $C_C$ atom electron deficient and highly reactive. Thus, facile attack of the the $O_{surface}$ atom on the $C_C$ occurs to generate the tetrahedral intermediate #2 in Figure 3(b).

It should be noted that we only consider the reaction pathway going through the intermediate steps ((a), (b), (c) in Figure 2) suggested from the previous work [23] of EC decomposition on $LiMn_2O_4$ [100] surface. As long as we find a reaction pathway which has very low reaction barrier, e.g. $E_{barrier}$=17meV for the bare surface case with $Li^+$ in the nearby environment (from the electrolyte), we can conclude that the EC ring-opening process is an extremely fast chemical reaction no matter if there are some other possible intermediate steps (or reaction pathways) for the EC ring-opening reaction.

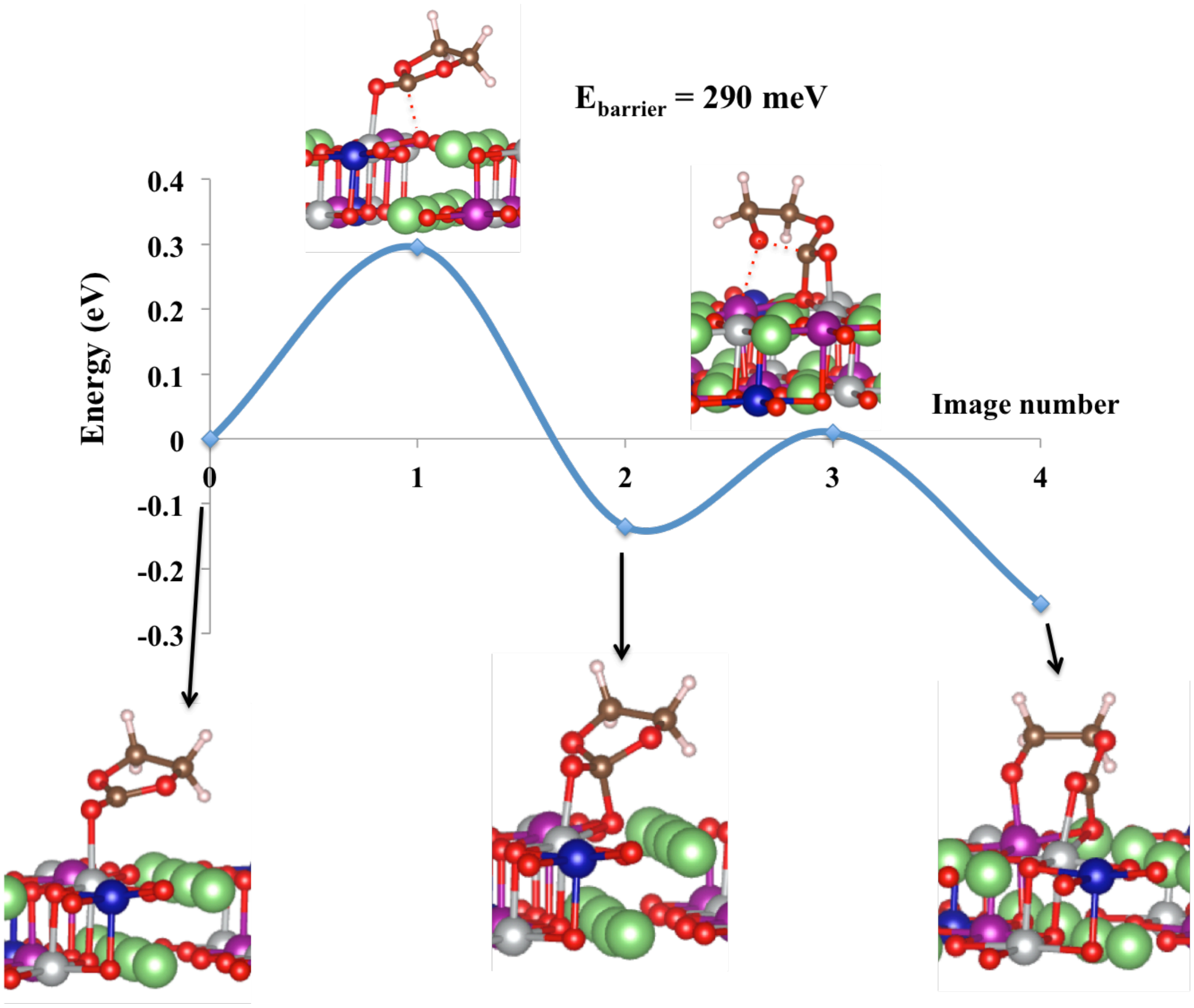

(a)

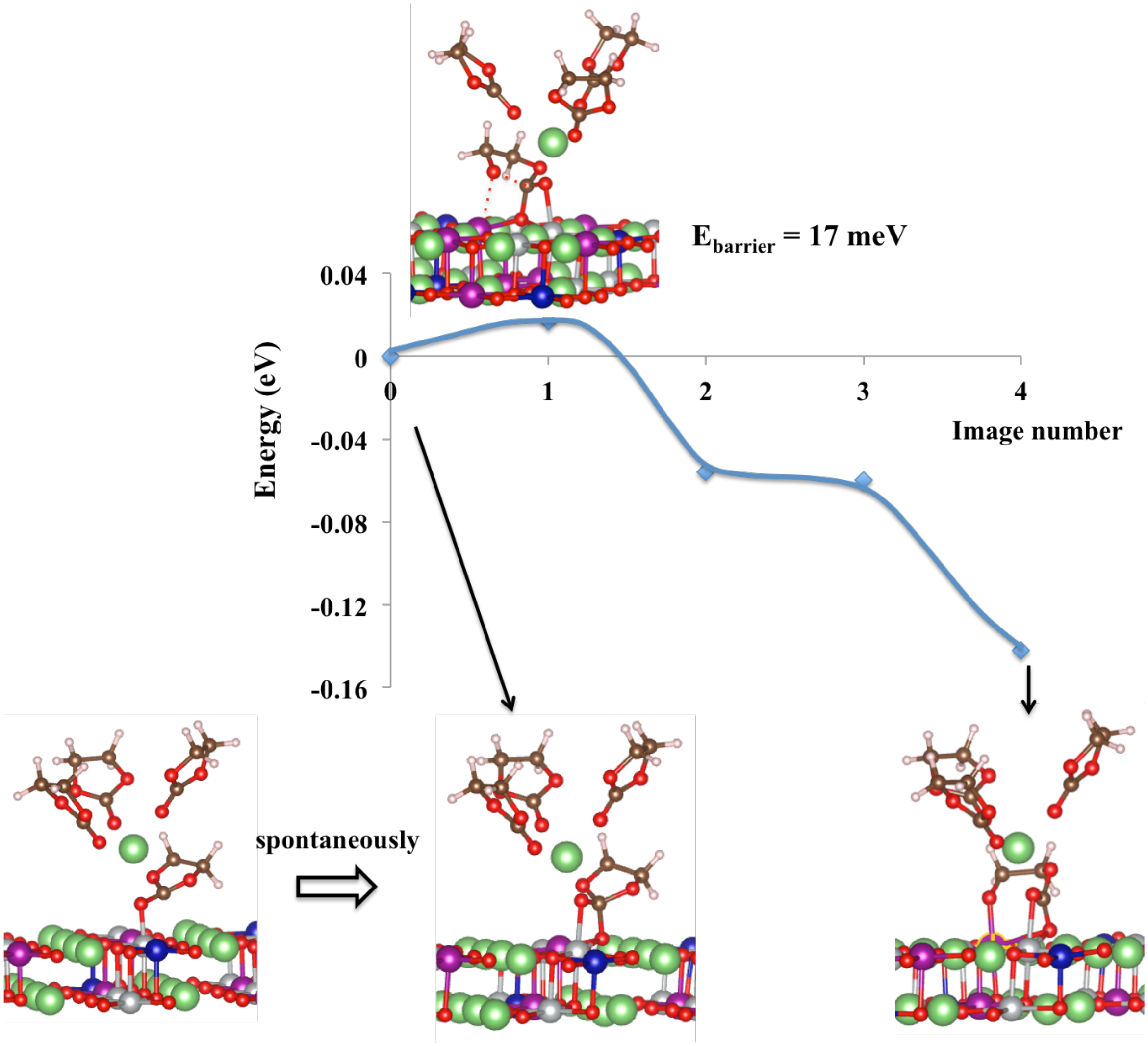

(b)

**Fig. 3**: Reaction energy profile with respect to the image numbers for the single EC molecule bond breaking process on (a) bare NMC surface (UHV conditions) and (b) bare NMC surface with a $Li^+$ ion and its coordinated EC molecules from the electrolyte.

### 3.4 Effect of surface termination on reaction energy barriers of the EC ring-opening process

In addition to the electrolyte environment at the surface, another important factor to consider is the cathode surface termination. Leung's previous simulation work studied the EC reaction only on the bare cathode surface.[23] Under more realistic conditions, especially when the cathode material is exposed to the air during synthesis or makes contact with the electrolyte solution during the battery assembly process, the cathode surface cannot be considered to be a completely bare surface. In this section, we consider a cathode surface terminated with a hydroxyl group –OH and a fluorine atom –F and we describe DFT calculations of the impact of these –OH and –F surface species on EC decomposition. As we have discussed in the Method Section 2.2, it is expected that coverage of both –OH and –F will be relatively low, which suggests that they will be relatively isolated on the surface. The atomistic configurations of the –OH and –F species on the NMC surface are shown in Fig. 4. The –OH group can form two possible configurations.[37] The first configuration consists of the H atom of –OH bonded to a surface oxygen atom (-OH type I in Fig. 4). The second configuration consists of the oxygen atom of the –OH group bonded to a surface transition metal atom (–OH type II in Fig. 4). We have calculated the energy barriers of the EC $C_C$-$O_E$ bond breaking process with –OH or –F terminations, within the presence of $Li^+$ ion and its coordinated EC molecules to take into account the influence of the electrolyte solution. The calculated energy barriers are shown in Table 1 and the plots of the reaction pathway configurations are shown in the SI Fig. S1-S3. We found that the –OH type II increases the reaction energy barrier significantly from 17 meV to 860 meV, and therefore has a stronger passivation effect on the cathode surface. In addition, while the –OH type I and –F also raise the reaction energy barrier compared to the bare surface case, the effect is significantly less dramatic. The reaction rates of EC decomposition for all the electrolyte environment barriers determined in this work are discussed in Section 4.

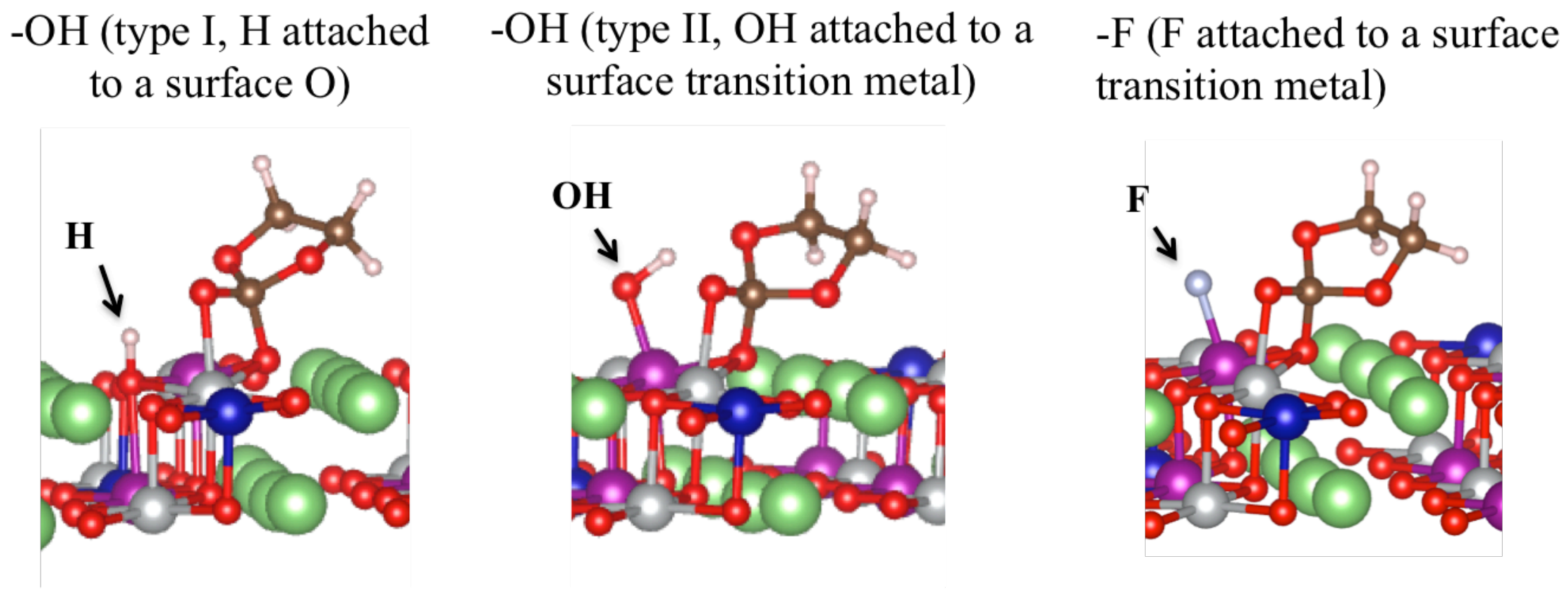

**Fig. 4**: Illustrations of the atomistic structures of the NMC surface with –OH (type I and II) and –F termination.

| Bare NMC | -OH type I ($O_{surface}$-H) | -OH type II ($TM_{surface}$-OH ) | $TM_{surface}$-F |
|---|---|---|---|
| 17 meV | 150 meV | 860 meV | 490 meV |

**Table 1**: Energy barriers of the EC (with the coordinated $Li^+$ from the electrolyte) $C_C$-$O_E$ bond breaking reaction on NMC surface with –OH and –F terminations.

## 4. Discussion and Implications

In this section, the reaction rates for EC $C_C$-$O_E$ bond breaking are calculated from an Arrhenius law, and the physical implications of these adsorbates on the overall EC decomposition process are discussed. The Arrhenius equation to estimate the reaction rate is shown as below:

$$R = k_0 \exp(-E_{barrier} / kT) \times C_R \times (1 - C_P), \qquad \text{Eq. 2}$$

where $R$ is the reaction rate per second per reaction site, the exponential prefactor $k_0$ is the typical molecular vibrational frequency and is estimated to be $10^{12}$/s at room temperature, $C_R$ is the reactant fraction (per site), $C_P$ is the product fraction (per site), and $E_{barrier}$ is the CI-NEB calculated energy barrier. Since we have focused on the initial steps of the EC reaction on the cathode surface when the electrolyte just makes contact with the cathode material, we have assumed $C_R$ = 1 and $C_P$ = 0 in Eq. 2. In Table 1, except for the –OH type II case ($TM_{surface}$-OH), all of our calculated energy barriers are ≤ 490 meV. If the reaction energy barrier is set to 490 meV, the reaction rate is as high as $6.5 \bullet 10^3$/s at room temperature ($T$ = 300K), which is equivalent to a time scale of 0.15 ms for a single EC decomposition reaction to happen. Therefore, although the –OH type I case ($O_{surface}$-H) and the –F case can passivate the cathode surface, the extent of this passivation is limited and the initial $C_C$-$O_E$ bond breaking step of EC decomposition still happens very quickly. We also note here that this fast ring-opening process is predicted by our model to be a purely chemical reaction (by saying “chemical” we mean that the EC molecules react with the cathode material which is not controlled by an externally set potential and that no such potential has been included in the reaction energy) rather than an electrochemical reaction, implying that the studied reaction rate is mostly independent of the voltage of the system, except for possible local field effects at the surface. Therefore, this EC $C_C$-$O_E$ bond breaking reaction would immediately occur once the NMC cathode particle surface contacts the electrolyte solution if the cathode has a bare surface or a surface partially terminated with –OH type I ($O_{surface}$-H) or –F.

We now consider the –OH type II case where an –OH group is bonded to the transition metal (Mn in our model) next to the Ni atom on the cathode surface. The calculated reaction barrier is 860 meV, the corresponding reaction rate is $3.6 \cdot 10^{-3}$/s, equivalent to a time scale of 232 s (~ 4 min) for a single EC decomposition reaction to happen. This result indicates that the –OH type II adsorption has a much stronger passivation effect on the cathode surface and significantly suppresses the $C_C$-$O_E$ bond breaking process required for EC decomposition. However, the predicted reaction rate even with –OH type II passivation is still very fast compared to what might be observed in a battery. To better understand the decomposition rates expected for a real battery, we consider the case of a Hybrid Electric Vehicle (HEV) battery and make the assumption that it decomposes at most 10% of its electrolyte over one year at T = 300 K (the desired operation temperature range for a HEV battery is from 15°C to 35°C (288K-308K) [44], and here we simply choose T=300K for our calculation). Assuming standard reaction rate models, this rate limit requires a rate-limiting step barrier larger than 1.03 eV (see SI Section 3 for the calculation details of this estimation). Therefore, even though the –OH type II species on the surface (under a relatively low surface coverage condition) is able to make the EC ring-opening process occur much slower, the passivation effect is still not strong enough to provide a long-term protection for the cathode surface.

A possible explanation for the stronger passivation effect of the –OH type II case is that the –OH group occupies the transition metal site (Mn in our model), which is available in the unpassivated surface for the $O_E$ in the EC molecule to bond with after the $C_C$-$O_E$ bond is broken. Since the Mn site is no longer available for the $O_E$ to bond to, the EC molecule has to go through another reaction pathway to break the $C_C$-$O_E$ bond and find another transition metal site on the surface to bond with the $O_E$. The corresponding atomistic configurations of the reaction intermediate states are shown in Fig. S2 in the SI. The two highest energy barrier cases shown in Table 1, i.e., the –OH type II (860 meV) and –F (490 meV) terminations proceed through this new reaction pathway, which is different from the pathway shown in Fig. 1 (a) and (b). This new reaction pathway is the origin of the significantly higher barriers compared to the other two cases, i.e., the bare surface (17 meV) and the –OH type I termination (150 meV). It should be noted that if the –OH and the –F coverage is high enough to cover almost all of the transition metal sites on the NMC surface, within the scope of discussion in this atomistic modeling work, all of our simulated reaction pathways would not apply to the EC ring opening process because the carbonyl oxygen $O_C$ and the ethylene oxygen $O_E$ would not be able to find an available transition metal site with which to bond. We suspect that under such conditions some other species from the electrolyte, e.g. $PF_5$ (one of the decomposed products from $LiPF_6$)[24], may react with the EC molecule. However, the above case (very high surface coverage of –OH and –F) is not very likely, for reasons discussed in Section 2.2.

Besides the different surface terminations, the lithiated state of the cathode material may also influence the reaction barrier of the EC bond-breaking process. First, the highly-delithiated cathode material may induce a faster proton transfer (from EC to cathode) compared to the bond-breaking process [42-43]. Second, the more delithiated state of the cathode material corresponds to a higher oxidation state of the system, for example, more high-valent $Ni^{4+}$/$Co^{4+}$/$Mn^{4+}$ transition metal atoms at the surface might enhance EC bond-breaking reaction rate because the high-valent transition metal are easier to bond with the $O_E$ atom in the EC molecule. Generally speaking, the more delithiated cathode material are expected to induce faster EC decomposition reactions. As our rates are already very fast, such a change does not alter any of our conclusions and implications in this work, which would also apply to a partially delithiated cathode.

Once the organic electrolyte molecule EC contacts with the NMC cathode surface, a fast reaction of EC $C_C$-$O_E$ bond breaking will happen on the cathode surface where no hydroxyl groups are present, leading to the formation of $C_C$-$O_E$ bond broken EC monolayer on the areas of initially bare surface. By considering the overall cathode surface, the above analysis indicates that the majority of the NMC cathode surface will be covered by a monolayer consisting of the hydroxyl groups (leftover from the battery cathode synthesis and construction) and the $C_C$-$O_E$ bond broken EC molecules within 1 millisecond after the organic electrolyte solution contacts with the cathode particles. Because this monolayer forms so quickly, this initial bond breaking process cannot be the rate-limiting step of the overall EC molecule decomposition reaction. This insight implies that true rate-limiting step is likely some further reaction that occurs on the monolayer consisting of the bond broken EC molecules and the adsorbed –OH molecules, which might be certain reactions facilitated by $PF_5$ or another EC molecule (formation of an EC dimer) as suggested in the work by Tebbe, et al.[14] Some possible final products can be LiF, $Li_2O$, $Li_2CO_3$ which have been discussed in previous works [1-2, 16] as components in the SEI layer.

**5. Conclusions**

In conclusion, our study indicates that after the organic electrolyte molecule EC contacts with the NMC cathode surface, a fast reaction of EC $C_C$-$O_E$ bond breaking will happen on the cathode surface. This reaction will lead to the formation of a monolayer of $C_C$-$O_E$ bond broken EC molecules and/or their further decomposition products and previously adsorbed –OH molecules on the cathode surface and contribute to the formation of the SEI. Our atomistic model shows that this fast reaction can occur in the absence of any interfacial electric double layer, and it is a purely chemical (not electrochemical) reaction that occurs immediately after the contact of the cathode material with the organic electrolyte solution independent of the cell voltage. Our results show that the –OH type I and –F surface terminations can slightly passivate the NMC cathode surface against the initial step of the EC decomposition reaction. The calculated reaction barriers of the -OH type I and -F surface terminations lead to millisecond-scale or faster chemical reactions on the NMC surface. Since our estimated rate-limiting $E_{barrier}$ for a typical HEV battery is > 1.03 eV, the EC $C_C$-$O_E$ bond breaking reaction on the cathode surface cannot be the rate-limiting step of a complete EC molecule decomposition reaction. Possible rate-limiting steps may be reactions of EC and certain decomposed products of $LiPF_6$ with the already decomposed organic products (e.g. $C_C$-$O_E$ bond broken EC molecules) deposited on the cathode surface. Future studies of these steps will provide further insight into the cathode surface chemistry after the initial chemical reactions between the cathode material and the organic electrolyte solution. These insights would help further develop the understanding of SEI formation at the cathode surfaces and provide guidance for the future design of Li-ion batteries that exhibit less capacity and voltage fade, thus providing longer battery lifetime.

**ACKNOWLEDGEMENTS**

The authors gratefully acknowledge funding from The Dow Chemical Company and helpful conversations with Mark Dreibelbis, Brian Goodfellow and Thomas Kuech. Computations in this work benefitted from the use of the Extreme Science and Engineering Discovery Environment (XSEDE), which is supported by National Science Foundation grant number OCI-1053575.

**Supporting Information**

Pathways of the EC bond breaking reactions on the NMC surface with –OH and –F terminations, energetic comparison with previous theoretical work[24], estimation of a rate-limiting reaction barrier value, input files for DFT calculations (in .zip).

## Table of Contents

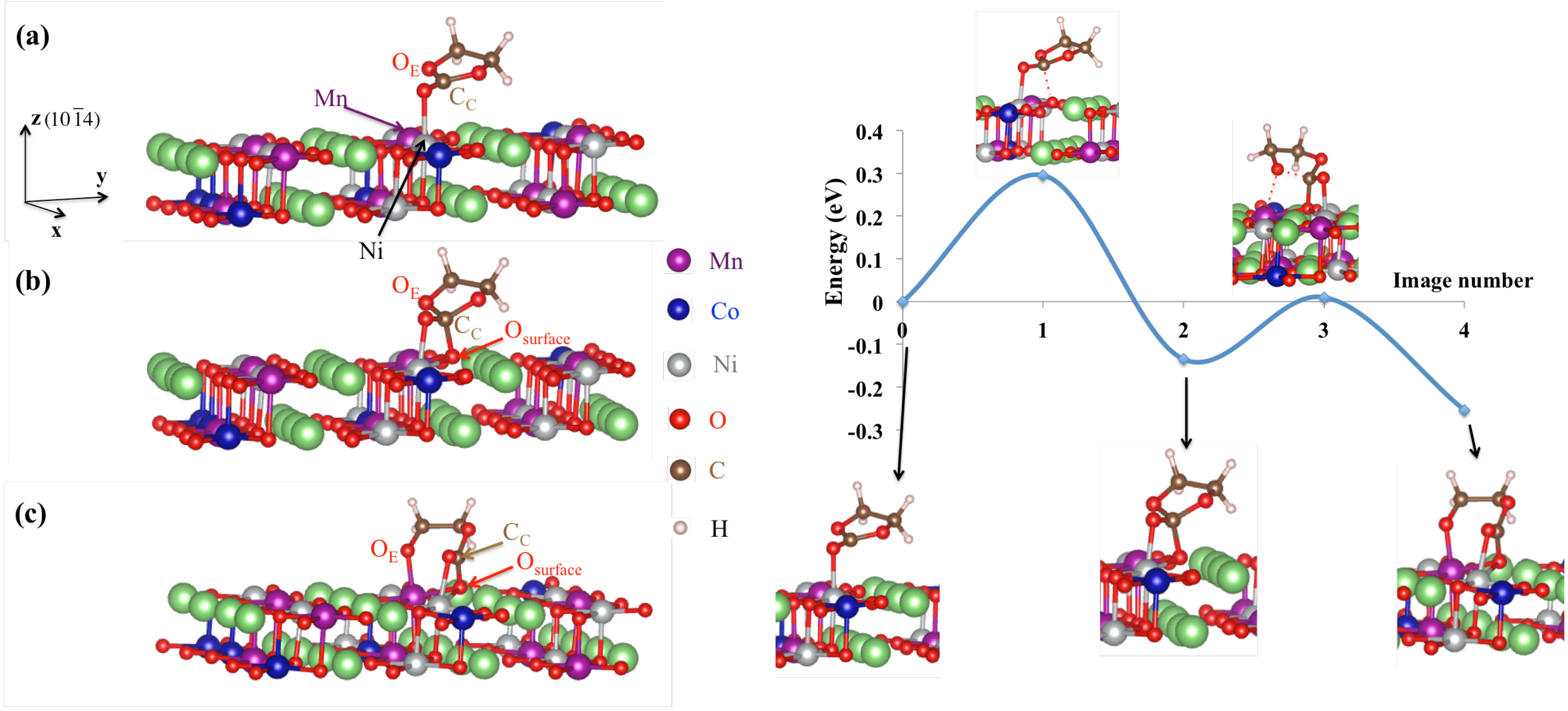